\documentclass[draftcls,onecolumn,twoside,journal]{IEEEtran}

\usepackage[Symbol]{upgreek}
\usepackage{cite}
\usepackage{graphicx}
\usepackage{subfigure}
\usepackage{amsmath}
\usepackage{array}
\usepackage{cases}
\usepackage{threeparttable}
\usepackage{amssymb}
\ifx\pdfoutput\undefined
\usepackage[hypertex]{hyperref}
\else
\usepackage[pdftex,hypertexnames=false]{hyperref}
\fi

\begin{document}
%
\title{Evaluate the Word Error Rate of Binary Block Codes with Square Radius Probability Density~Function}

\author{Xiaogang~Chen,~\IEEEmembership{Student Member,~IEEE,}
        Hongwen~Yang,~\IEEEmembership{Member,~IEEE,}\\
        Jian~Gu and Hongkui~Yang
\thanks{The authors are with the School of Telecommunication Engineering, Beijing University of Posts and Telecommunications,
        Beijing, 100876, China. (e-mail: debug3000@gmail.com; yanghongwen@263.net.cn)}}



\maketitle

\begin{abstract}
    The word error rate (WER) of soft-decision-decoded binary block
    codes rarely has closed-form. Bounding techniques are widely used
    to evaluate the performance of maximum-likelihood decoding algorithm.
    But the existing bounds are not tight enough especially for low signal-to-noise ratios and become looser when a
    suboptimum decoding algorithm is used.
    This paper proposes a new concept named \textit{square radius} probability
    density function (SR-PDF) of decision region to evaluate the WER. Based
    on the SR-PDF, The WER of binary block codes can be calculated precisely for ML and suboptimum decoders.
    Furthermore, for a long binary block code, SR-PDF can be approximated by \textit{Gamma} distribution
    with only two parameters that can be measured easily.
    Using this property, two closed-form approximative expressions are proposed
    which are very close to the simulation results of the WER of interesting.
\end{abstract}


\IEEEpeerreviewmaketitle

\section{Introduction}
    \PARstart{T}{he} performance evaluation of binary block codes with
    soft-decision-decoding in additive white Gaussian noise (AWGN) and fading channels has long
    been a problem in coding theory and practice. A closed-form
    expression of word error rate (WER) for popularly used long codes hasn't been derived as yet. Thus, bounding
    techniques are widely used for performance evaluation of maximum-likelihood (ML) decoding. The
    most popular upper bound is the \textit{union bound}. When the
    \textit{weight enumerating function} of a code is known, union bound
    presents a tight upper bound at signal-to-noise ratios (SNRs) above the cutoff rate limit but becomes useless at
    SNRs below the cutoff rate limit \cite{1}. Based on \textit{Gallager's first bounding
    technique} \cite{2}, some tighter upper bounds are presented
    \cite{3}\cite{4}\cite{5}. These bounds are tighter relative to union
    bound. But even Poltyrev's \textit{tangential sphere bound}
    \cite{6}, which has been believed as the tightest bound for binary
    block codes, still has a gap to the real value of WER
    \cite{7}\cite{8}\cite{9}. Additionally, these bounds are all based on
    ML decoding, which is too complex to implement in practice.
    When a suboptimum decoder is used, the WER will change but these bounds still keep their original value.

    In this paper, a new concept named \textit{square radius} probability density function (SR-PDF) of decision region is proposed
    to calculate the WER precisely at SNRs of interesting and any decoding algorithm.
    The basic premise is that in AWGN channel, when the encoder and decoding algorithm are
    given, the decision region is fixed, and WER is completely determined by
    the decision region. The SR-PDF proposed in this paper is unique for every encoder-channel-decoder models, thus any
    changes of channel and decoding algorithm can be reflected by it.
    Furthermore, when the codeword length $N\rightarrow\infty$, the asymptotic WER is exactly the complementary cumulative
    distribution probability of the normalized square radius and can be roughly characterized by the maximum point of the SR-PDF.
    Moreover, for popular long codes such as Turbo codes, Low-density parity check (LDPC) codes and Convolutional codes, etc,
    their SR-PDFs are close to \textit{Gamma} distribution,
    which implies that the exhausting measurement of SR-PDF is not needed, and
    only the mean and variance of the square radius are enough to get an approximated SR-PDF.
    Based on these properties, two closed-form approximative expressions of WER are proposed and
    their approximation errors are around the order of 0.1dB and 0.3dB respectively for WER above $10^{-3}$.

    The rest of this paper is organized as follows: Section II introduces the system model and the concept of
    SR-PDF of decision region, as well as the method of measuring the pdf. In section III, the WER in AWGN channel is derived using SR-PDF
    and several examples are followed to prove the validity of the proposed method. Section IV illustrates some important asymptotic properties
    of the relation of WER to the normalized SR-PDF.
    In section V, two closed-form approximative expressions are derived for WER calculation. Section VI extending the SR-PDF method to flat fading channel.
    The conclusion is drawn in section VII, including some discussion and remarks on the SR-PDF method and other applications.

\section{Preliminaries}
    \subsection{System Model}
        The following system model is used in the paper. Binary
        information bits are first encoded to binary block code. The code rate is $R=k/N$, where $N$ is the
        codeword length and $k$ is the information length. Then encoded bits are BPSK
        modulated (the resulting signal set is ${\rm {\bf S}}$), and transmitted in AWGN channel.
        The received signal is
        \begin{equation}
        \label{eq1}
        {\rm {\bf y}}={\rm {\bf s}}+{\rm {\bf n}}
        \end{equation}
        where ${\rm {\bf s}}=\left[ {s_1 ,s_2 ,...,s_N } \right]^T\in {\rm {\bf S}}$,
        $s_i \in \{ \pm \sqrt{E_s}=\pm\sqrt{RE_b}\}$, $E_s$ and $E_b$ are, respectively,
        the energy per code bit and the energy per information bit.
        ${\rm {\bf n}}=\left[ {n_1 ,n_2 ,...n_N } \right]^T$ is the additive white Gaussian noise with zero-mean and variance $\frac{N_0 }{2}$.
        In this paper, SNR (signal to noise power ratio) is defined as
        $\beta =\frac{2E_s}{N_0}$. It is also assumed that the signal $\bf s$ is detected coherently and decoded
        with some algorithm at the receiver, and the channel side information is known if it is required.

    \subsection{Decision Region and Square Radius Probability Density Function}
\begin{figure}[htbp]
            \begin{center}
                \setlength{\abovecaptionskip}{0pt}
                \includegraphics[width=0.5\textwidth,keepaspectratio=true]{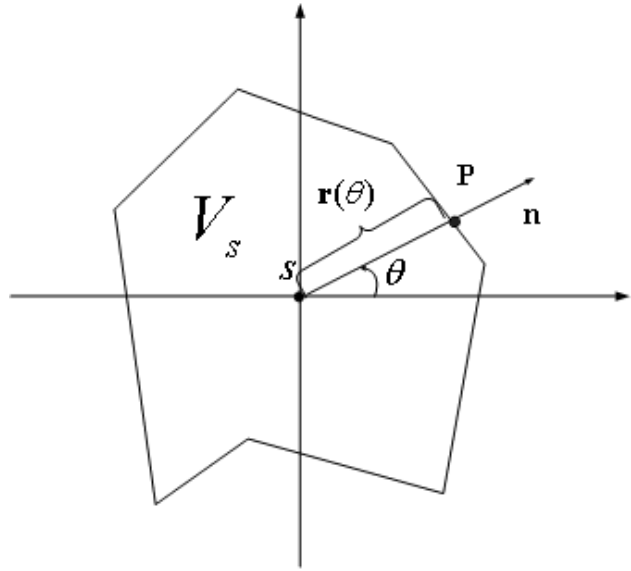}
                \caption{Decision region of ${\rm {\bf s}}$ and its radius for two dimensional codes.}
                \label{f1}
            \end{center}
        \end{figure}

        The decision region $V_s $ of signal ${\rm {\bf s}}$ is a set in $N$ dimensional
        Euclidean space ${\bf R}^N$. Consider the decoder as a function that maps the received vector $\rm {\bf y}$
        to a transmitted signal $\rm{\bf s} \in \rm {\bf S}$, $f_{decoder}: {\bf R}^N\rightarrow \bf S$, then the decision region can be defined as
        $V_s=\left\{ {\bf y}|{\bf y} \in {\bf R}^N,f_{decoder}(\rm{\bf y})={\rm{\bf s}}
        \right\}$, i.e. the domain of $\bf s$. Whenever the decoder is specified, the decision region is fixed.
        When the received vector ${\rm {\bf y}}$ lies inside $V_s $,
        it will be correctly decoded to ${\rm {\bf s}}$, otherwise a decoding error occurs. For linear block codes investigated in this paper,
        all the codewords have the same decision region. Fig.~\ref{f1} is an example of two dimensional decision region, where ${\rm {\bf n}}$ is the additive noise,
        ${\rm {\bf \theta }}$ is the direction of ${\rm {\bf n}}$. Point ${\rm {\bf p}}$ is on the
        boundary of decision region and the radial originated from
        ${\rm {\bf s}}$ along ${\rm {\bf \theta }}$. Connecting ${\rm {\bf
        s}}$ and ${\rm {\bf p}}$, the length of vector ${\rm {\bf r}}({\rm {\bf \theta }})$ is the radius
        of the decision region along the direction of ${\rm {\bf n}}$.
        Define $l({\rm {\bf \theta }})=\left| {{\rm {\bf r}}({\rm {\bf \theta }})} \right|^2$
        as the \textit{square radius} in the direction of $\theta$. Because $\bf n$ is a random variable,
        so $l(\theta)$ is a random variable. The pdf of $l({\rm {\bf \theta }})$ is denoted as $p_l (l)$ and is abbreviated as SR-PDF.
        This concept can be extended to $N$ dimensional space, where $\uptheta$ is determined by
        $N-1$ angles (e.g. azimuth and elevation in three dimensional space).

        The SR-PDF is too complex to work out in analytical way for long codes, but it can be measured with simulation:
        For the system model above, generate a white Gaussian noise vector ${\rm {\bf n}}=(n_1 ,n_2 ,...n_N )$,
        normalize $\bf {n}$ to ${\rm {\bf {n}'}}=\frac{{\rm {\bf n}}}{\left| {\rm {\bf n}} \right|}=(1 ,{\rm {\bf
        \uptheta }})$, where ${\rm {\bf \uptheta }}$ is the direction of ${\rm {\bf n}}$,
        scale ${\rm {\bf {n}'}}$ by $\lambda$ and send the vector ${\rm {\bf y}}={\rm {\bf s}}+\lambda{\rm {\bf {n}'}}$ to the decoder.
        There exists a $\hat \lambda>0$ such that $f_{decoder}({\bf s}+\lambda{\bf n}')={\bf s},~\forall\lambda\le\hat\lambda$
        and $f_{decoder}(\bf{s}+\lambda\bf{n}') \neq {\bf s},~\forall \lambda>\hat\lambda~$.
        In this paper, we will assume that the decision region is simply connected\footnote{Note that when iterative decoding algorithm is used for Turbo and LDPC decoding, as the iteration increases,
        it is possible, particularly on very noisy channel,
        for the decoder to converge to the correct decision and then diverge again \cite{10}.
        i.e. it is possible that ${\bf s}+\lambda_1{\bf n}'$ can be decoded in error even if $f_{decoder}\left({\bf s}+\lambda_2\bf{n}'\right)=\bf{s}$,
        where $\lambda_1<\lambda_2$. This implies that the decision region of iterative decoder may not be a
        \emph{simply connected region}, but the probability of this exception is much smaller than the WER.}.
        Consequently, $|{\rm {\bf r}}({\rm {\bf \uptheta }})|=\hat{\lambda}$ is the radius in direction of $\bf n$,
        and $l(\uptheta)=\hat{\lambda}^2$ is the square radius. Note that $\bf r(\uptheta)$ is related to the code bit energy and codeword length.
        This effect can be removed by normalizing the square radius by $NE_s$, i.e. $l_n=\frac{l}{NE_s}$.
        The resulting SR-PDF is referred to as \textit{normalized SR-PDF} and is denoted as $p_{l_n }(l_n)$ hereafter.
        With sufficient tests, the approximated normalized SR-PDF can be obtained.

\begin{figure*}[htbp]
            \begin{center}
            \includegraphics[width=0.5\textwidth,keepaspectratio=true]{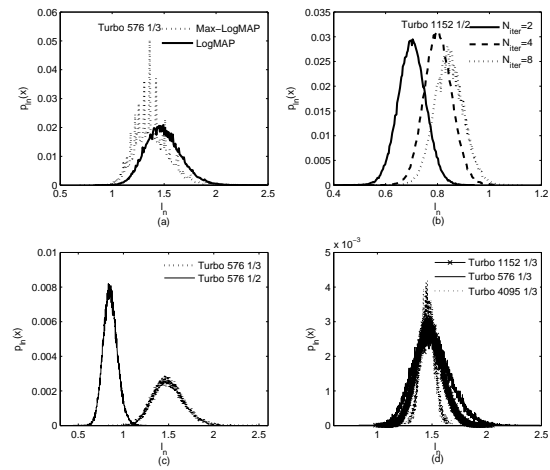}
            \caption{Normalized SR-PDF of Turbo codes for different (a) decoding algorithm;
            (b) iterations; (c) code rate; (d) codeword length. ``Turbo 1152 1/3'' stands for a turbo code
            with code rate 1/3 and codeword length 1152.}
            \label{f2}
            \end{center}
        \end{figure*}

        Fig.~\ref{f2} is the normalized SR-PDF of Turbo codes with different codeword lengths,
        code rates and decoding algorithms in AWGN channel. Turbo codes used in this paper is defined in \cite{13}.
        Obviously, all the changes in the encoder and decoder
        will be reflected by the SR-PDF.
        Larger decision region implies that the code can tolerate larger noise,
        therefore, the SR-PDF which occurs in the righter side will have a better WER performance.
        Subsequent sections will mainly discuss the relation between SR-PDF and WER.

\section{Word Error Rate In AWGN Channel }
    In AWGN channel, the decision region and the corresponding SR-PDF is completely determined whenever the decoder is specified,
    no matter whether it is a maximum likelihood (ML) decoder like the Viterbi decoding for Convolutional codes or a suboptimal decoder such as
    the iterative decoders for Turbo and LDPC codes.
    For a binary linear block code, all the codewords have the same error rate. Therefore, without loss of generality,
    assume a codeword $\bf{s}\in\bf{S}$ is transmitted. The word error rate is the probability that
    the received vector ${\bf y} \notin V_s$ conditioned on ${\bf s}$, that is

    \begin{equation}
    \label{eq2}
    \begin{split}
    P_e &=1-\oint\limits_{V_s } {p\left( {{\rm {\bf y}}\vert {\rm {\bf s}}}
    \right)d{\rm {\bf y}}} \\
    &=1-\oint\limits_{V_s-s } {p\left( {{\rm {\bf y}}-{\rm {\bf s}}} \right)d{\rm
    {\bf y}}} \\
    &=1-\oint\limits_{V_s^{'} } {p\left( {\rm {\bf n}} \right)d{\rm {\bf n}}} \\
    \end{split}
    \end{equation}
    where $V_s^{'}=V_s-s$ is the shift of the decision region from $s$ to origin. Denoting ${\rm {\bf n}}$ in polar coordinates,
    (\ref{eq2}) becomes
    \begin{equation}
    \label{eq3}
    \begin{split}
    P_e &= 1-\oint\limits_{V_s^{'} } {p_{\rho ,{\rm {\bf \uptheta }}} \left( {\rho
        ,{\rm {\bf \uptheta }}} \right)d\rho d{\rm {\bf \uptheta }}} \\
        &= 1-\oint\limits_{V_s^{'} } {p_{\rho \vert {\rm {\bf \uptheta }}} \left( \rho
        \right)p_{\rm {\bf \uptheta }} ({\rm {\bf \uptheta }})d\rho d{\rm {\bf \uptheta
        }}} \\
        &= 1-\int_{\rm {\bf \uptheta }} {\left[ {\int_{\rho \le |r({\rm {\bf \uptheta }})|}
        {p_{\rho \vert {\rm {\bf \uptheta }}} \left( \rho \right)d\rho } }
        \right]p_{\rm {\bf \uptheta }} \left( {\rm {\bf \uptheta }} \right)d{\rm {\bf
        \uptheta }}} \\
        &= \int_{\rm {\bf \uptheta }} {\left[ {\int_{\rho >|r({\rm {\bf \uptheta }})|}
        {p_{\rho \vert {\rm {\bf \uptheta }}} \left( \rho \right)d\rho } }
        \right]p_{\rm {\bf \uptheta }} \left( {\rm {\bf \uptheta }} \right)d{\rm {\bf
        \uptheta }}} \\
        &= E_{\rm {\bf \uptheta }} \left[ {\int_{\rho >|r({\rm {\bf \uptheta }})|} {p_{\rho
        \vert {\rm {\bf \uptheta }}} \left( \rho \right)d\rho } } \right] \\
    \end{split}
    \end{equation}
    Note that the integral in the bracket is the decoding error probability
    conditioned on the direction of a noise realization, so
    \begin{equation}
    \setlength\jot{12pt}
    \label{eq4}
    \begin{split}
    P_{e\vert {\rm {\bf \uptheta }}} &=\int_{\rho >|r({\rm {\bf \uptheta }})|}
    {p_{\rho \vert {\rm {\bf \uptheta }}} \left( \rho \right)d\rho } \\
    &=P\left[ {\rho \ge \left| {r({\rm {\bf \uptheta }})} \right|} \right] \\
    &=P\left[ {\rho ^2\ge l({\rm {\bf \uptheta }})} \right] \\
    \end{split}
    \end{equation}
    Expectation over ${\rm {\bf \uptheta }}$ is equivalent to expectation over $l$, thus the average error probability is
    \begin{equation}
    \setlength\jot{9pt}
    \label{eq5}
    \begin{split}
    P_e &=E_\uptheta \left[ {P_{e\vert {\rm {\bf \uptheta }}} } \right] =E_{l} \left[ {P\left( {\rho ^2\ge l } \right)} \right] \\
    &=\int_0^\infty {P\left( {\rho ^2\ge x } \right)p_{l} \left( x \right)dx} \\
    &=\int_0^\infty {P\left( {\rho ^2\ge NE_sx } \right)p_{l_n} \left( x \right)dx} \\
    \end{split}
    \end{equation}
    where $\rho ^2=\left| {\rm {\bf n}}\right|^2=\sum\limits_{i=1}^N {n_i^2 } $. Define $x=\rho ^2$,
    then $x$ is chi-square distributed \cite{11} with $N$ degrees of freedom. The pdf of $x$ is
    \begin{equation}
    \label{eq6}
    p(x)=\frac{1}{N_0 ^{N/2}\Gamma \left( {N \mathord{\left/ {\vphantom {N 2}}
    \right. \kern-\nulldelimiterspace} 2} \right)}x^{N \mathord{\left/
    {\vphantom {N 2}} \right. \kern-\nulldelimiterspace} 2-1}e^{-x
    \mathord{\left/ {\vphantom {x {N_0 }}} \right. \kern-\nulldelimiterspace}
    {N_0 }},\mbox{ }x\ge 0
    \end{equation}
    where $\Gamma (x)$ is the \textit{gamma function} \cite{12}. When $x$ is an integer and $x>0$, $\Gamma
    (x+1)=x!$ (the factorial of $x$). So
    \begin{equation}
    \label{eq7}
    \setlength\jot{10pt}
    \begin{split}
    P_{e\vert {\rm {\bf \uptheta }}} &=P\left[ {\rho ^2\ge l} \right] \\
    &=1-\int_0^{NE_sl_n} {\frac{1}{N_0 ^{N/2}\Gamma \left( {N \mathord{\left/ {\vphantom {N
    2}} \right. \kern-\nulldelimiterspace} 2} \right)}x^{N/2-1}e^{-x/N_0 }} dx \\
    &=1-\frac{1}{\Gamma \left( {N \mathord{\left/ {\vphantom {N 2}} \right.
    \kern-\nulldelimiterspace} 2} \right)}\int_0^{N\beta l_n/2}
    {e^{-x}x^{\frac{N}{2}-1}dx} \\
    &=1-\frac{1}{\Gamma \left( {N \mathord{\left/ {\vphantom {N 2}} \right.
    \kern-\nulldelimiterspace} 2} \right)}\gamma \left( {\frac{N}{2},\frac{N\beta l_n}{2}} \right) \\
    \end{split}
    \end{equation}
    where $\gamma (\alpha ,x)$ is the \textit{incomplete gamma function} defined as \cite{12}
    \begin{equation}
    \label{eq8}
    \gamma (\alpha ,x)=\int_0^x {e^{-t}t^{\alpha -1}dt}, \quad \alpha>0
    \end{equation}
    when $\alpha $ is an integer \cite{12},
    \begin{equation}
    \label{eq9}
    \gamma (\alpha ,x)=(\alpha-1 )!\left( 1-e^{-x}\sum\limits_{m=0}^{\alpha -1} {\frac{x^m}{m!}}\right)
    \end{equation}
    substitute (\ref{eq7}) into (\ref{eq5}),
    \begin{equation}
    \label{eq10}
    P_e =E_\uptheta\left[P_{e|\uptheta}\right]=E_l\left[P\left(\rho^2>l\right)\right]
    =\int_0^\infty {\left[ {1-\frac{1}{\Gamma \left( {N \mathord{\left/
    {\vphantom {N 2}} \right. \kern-\nulldelimiterspace} 2} \right)}\gamma
    \left( {\frac{N}{2},\frac{N\beta x}{2}} \right)} \right]p_{l_n} (x)} dx
    \end{equation}
    (\ref{eq10}) show that, WER of linear binary codes under AWGN channel is completely  determined by the normalized
    SR-PDF through a one dimensional integral.

\begin{figure*}[htbp]
    \centerline{
    \subfigure[]{\includegraphics[width=0.5\textwidth,keepaspectratio=true]{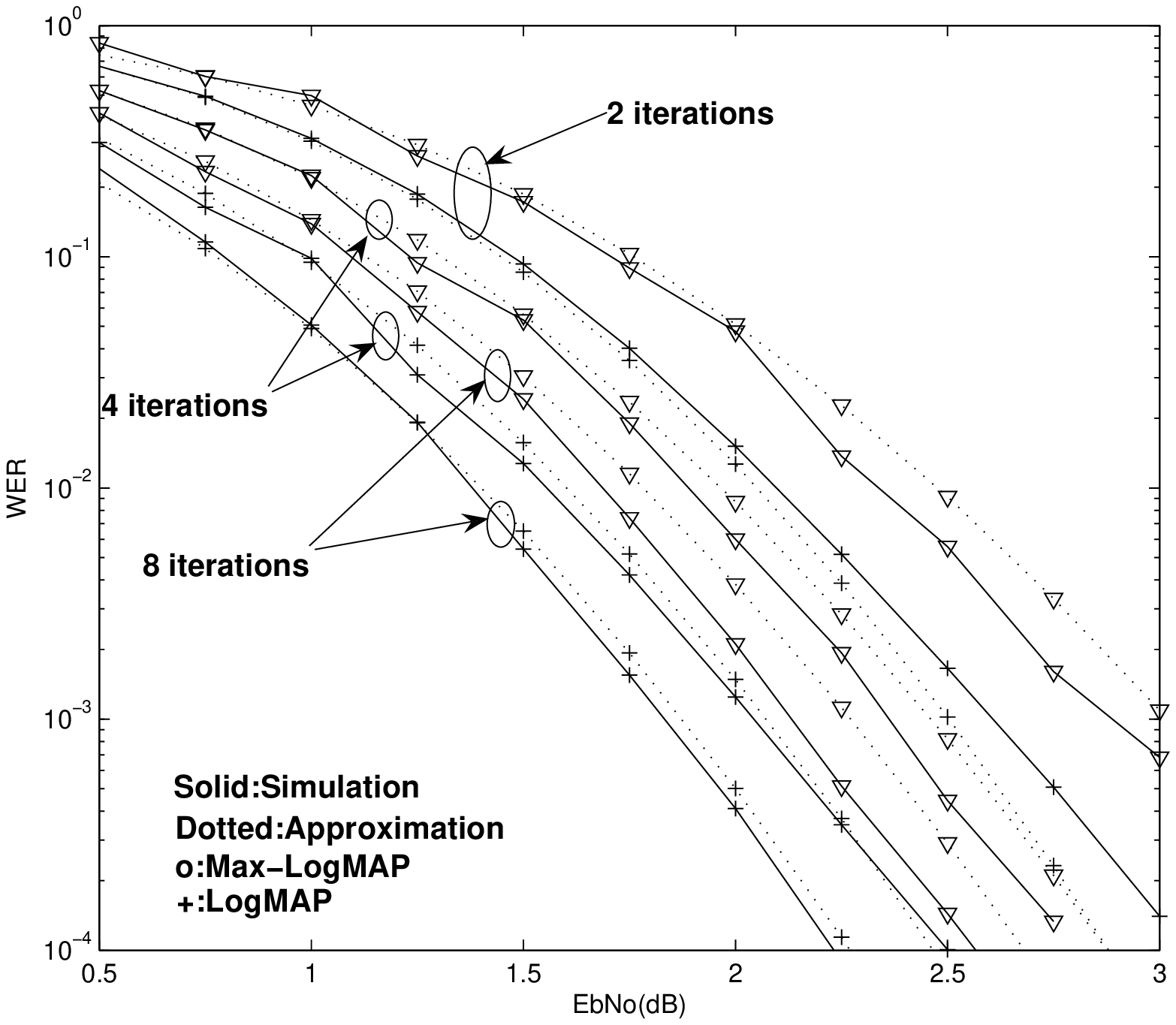} \label{f3a}}
    \hfil
    \subfigure[]{\includegraphics[width=0.5\textwidth,keepaspectratio=true]{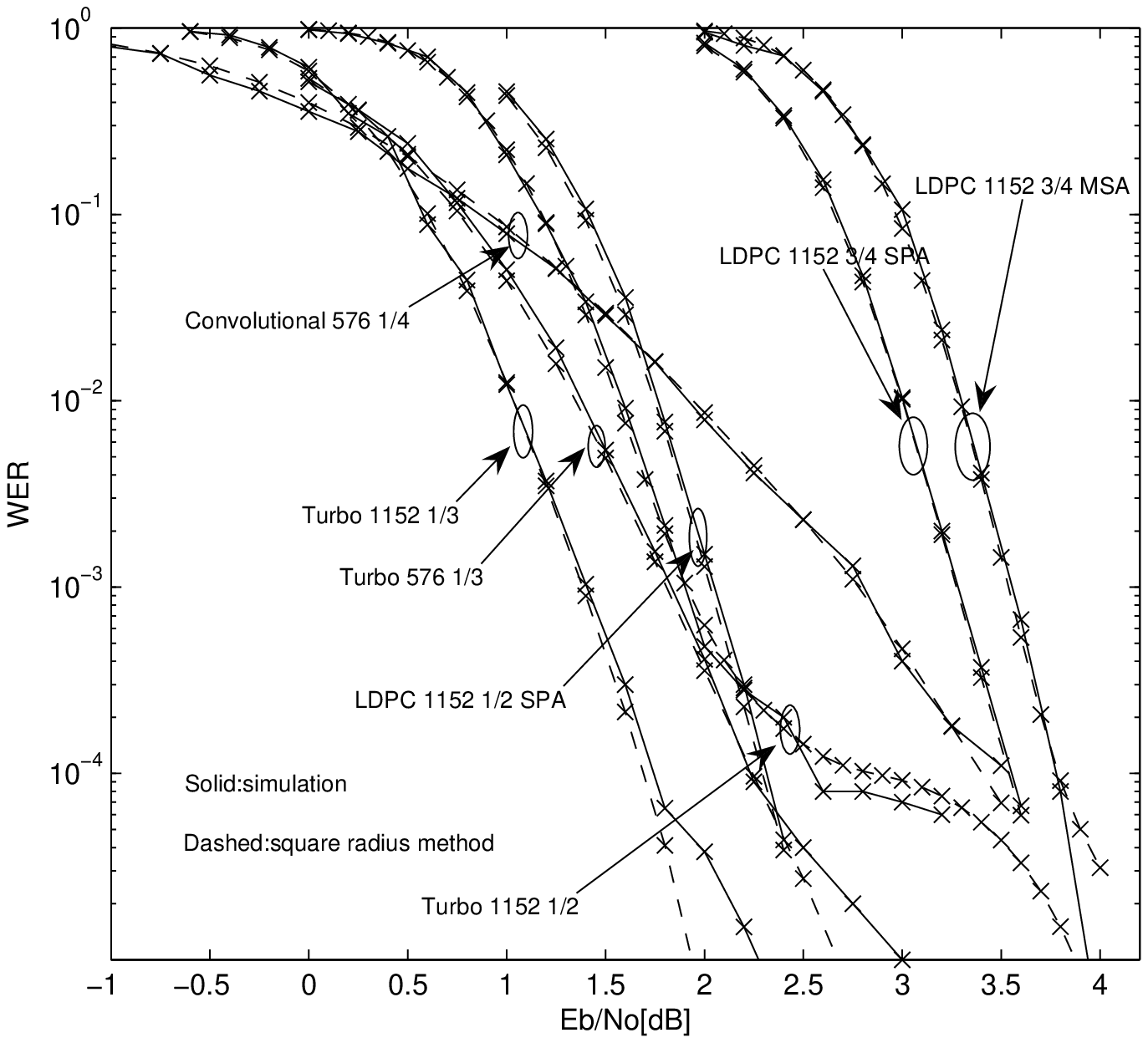} \label{f3b}}}
    \centering
    \caption{Comparison of the simulated WER and the WER evaluated with (\ref{eq10}) in AWGN channel.
    (a)The same Turbo codes with different maximum iterations. $N=576$, $R=1/3$,
    decoding algorithm is Log-MAP and Max-Log-MAP; (b)Different codes, including Convolutional codes,
    Turbo codes and LDPC codes.}
    \label{f3}
\end{figure*}

    Simulations are used to verify (\ref{eq10}). Three error control codes commonly used in wireless
    communications are considered, including Convolutional codes \cite{12}, Turbo codes \cite{12}
    and LDPC codes \cite{14}. Fig.~\ref{f3} is the comparison of WER between simulation and
    that evaluated from (\ref{eq10}). For the simulation, each point on the curve is obtained by $10^6$ tests;
    For the results of (\ref{eq10}), $10^5$ radius are measured to
    get $p_{l_n} (l_n)$, which is then substituted into (\ref{eq10}) to get the WER.
    Fig.~\ref{f3a} shows the WER of Turbo code for different maximum iterations. The parameter is $N=576$, $R=1/3$
    and the decoding algorithms are Log-MAP and Max-log-MAP.
    Fig.~\ref{f3b} are the WERs of Turbo, LDPC and
    Convolutional codes with different code lengths and code rates.
    The decoding algorithms are log-MAP with 8 maximum iterations for turbo code, soft Viterbi algorithm for Convolutional code,
    sum-product algorithm (SPA) and min-sum algorithm (MSA) with 25 maximum iterations and layered decoding for LDPC codes.
    It can be seen from these figures that the WER evaluated from (\ref{eq10}) matches very well with the simulation results
    except for large SNRs. The mismatch in large SNR region maybe caused by the inaccurateness of the simulation,
    or the inaccurateness of the
    ``left tail'' (this will be further explained in section IV) of $p_{l_n} (l_n)$, both of which are difficult to be measured precisely
    owning to infinitesimal probability. Moreover, the SR-PDF method can trace the change of decoder,
    e.g. the number of iterations, any modifications of the algorithm and etc. while the bounds such as
    union bound and so on cannot do this.

\section{Asymptotic Properties of WER}
    Eq.(\ref{eq10}) shows that the WER of a binary linear code is completely determined by the normalized SR-PDF, $p_{l_n}(x)$.
    When the codeword length, $N\rightarrow\infty$, there are some asymptotic properties of WER which will be discussed in this section.
    \newtheorem{theorem}{Property}
    \begin{theorem}
        Define $\tau=1/\beta$ and $P_e^a$ as the WER when codeword length $N\rightarrow\infty$, i.e.
        $P_e^a(\tau)=\lim\limits_{N\rightarrow\infty}P_e(N,\beta)$ where $P_e(N,\beta)$ is defined by
        (\ref{eq10}). Then, $P_e^a(\tau)$ is exactly the cumulative distribution function (CDF)
         of the normalized square radius:
        \begin{equation}
          \label{eq2261}
            P_e^a(\tau)=\lim_{N\rightarrow\infty }P_e(N,\beta)=\int_0^{1/\beta}p_{l_n}(x)dx=\int_0^{\tau} p_{l_n}(x)dx=F_{l_n}(\tau),
        \end{equation}
            where $F_{l_n}(x)$ is the CDF of normalized square-radius
            $l_n$.
    \end{theorem}
    \begin{proof}
    Define
    \begin{equation}
    \label{a1}
        f(\alpha,t) \triangleq {\frac{1}{\Gamma(\alpha)}}\gamma(\alpha,\alpha t)=\frac{\int_0^{\alpha t}e^{-x}x^{\alpha-1}dx}{\Gamma(\alpha)},
        \quad t\geq0,~\alpha>0
    \end{equation}
    It is obvious that $f(\alpha,0)=0$ and $f(\alpha,\infty)=1$.
    Taking derivative with respect to $t$,
    \begin{equation}
      \label{a4}
      \frac{\partial f}{\partial t}=\frac{\alpha e^{-\alpha t}(\alpha t)^{\alpha-1}}{\Gamma(\alpha)},\quad t\geq0
    \end{equation}
    For a large $\alpha$, $\Gamma(\alpha)$ can be approximated with Stirling's Series \cite{12}
    \begin{equation}
      \label{a2}
      \Gamma(\alpha)=e^{-\alpha} \alpha^{\alpha-\frac{1}{2}}\sqrt{2\pi}\left(1+\frac{1}{12\alpha}+\frac{1}{288\alpha^2}-\cdots\right)
      \approx e^{-\alpha} \alpha^{\alpha-\frac{1}{2}}\sqrt{2\pi}
    \end{equation}
    Substitute into (\ref{a4})
    \begin{equation}
      \label{a3}
      \frac{\partial f}{\partial t}
      =\frac{\alpha e^{-\alpha t}(\alpha t)^{\alpha-1}}{e^{-\alpha} \alpha^{\alpha-\frac{1}{2}}\sqrt{2\pi}}
      =\sqrt{\frac{\alpha}{2\pi t^2}}\left(te^{1-t}\right)^\alpha,\quad t\geq0
    \end{equation}
    The term $te^{1-t}$ is always less than 1 when $t\neq 1$. Thus,
    \begin{equation}
      \label{a5}
      \mathop {\lim }\limits_{\alpha \to \infty }\frac{\partial f}{\partial t}=
      \begin{cases}
        0\quad\  t\neq1\\
        \infty \quad t=1
      \end{cases}
    \end{equation}
    $f(\alpha,t)$ is a continuous function of $t$. Therefore, as $\alpha$ approaches infinity, $f(\alpha,t)$ approaches to
    a unit step function:
    \begin{equation}
      \label{a6}
      \mathop {\lim }\limits_{\alpha \to \infty }{\frac{1}{\Gamma(\alpha)}}\gamma(\alpha,\alpha t)=
      \begin{cases}
        f(\alpha,\infty)=1\quad t>1\\
        f(\alpha,0)=0\quad \ 0\leq t<1
      \end{cases}
    \end{equation}
    Applying (\ref{a6}) to (\ref{eq10}) will obtain (\ref{eq2261}).
    \end{proof}

    This property can also be explained with \textit{the law of large numbers}:
    Dividing $\rho ^2$ and $l\left( {\rm {\bf \uptheta }} \right)$ in (\ref{eq7}) by codeword length $N$
    \begin{equation}
    \label{eq14}
        P_{e\vert {\rm {\bf \uptheta }}} =P\left[ {\frac{\rho ^2}{N}>\frac{l({\rm {\bf \uptheta }})}{N}} \right]
    \end{equation}
    because $\rho^2=\sum\limits_{i=1}^N {n_i^2 }$, where $n_i^2 ,i=1,2,...N$ are independent and identically distributed (i.i.d)
    variables and $E(n_i^2 )=\frac{N_0}{2} $, based on the law of large numbers
    \begin{equation}
    \label{eq15}
    \mathop {\lim }\limits_{N\to \infty } P\left\{ {\left|
    {\frac{\sum\limits_{i=1}^N {n_i^2 } }{N}-\frac{N_0 }{2}} \right|=\left|{\frac{\rho ^2}{N}-\frac{E_s}{\beta }} \right|<\varepsilon } \right\}=1
    \end{equation}
    where $\varepsilon $ is a positive number arbitrarily small. (\ref{eq15}) implies that $\frac{\rho^2}{N}$
    approaches a constant, $\frac{E_s}{\beta}$, as $N$ approaches infinity.
    Based on (\ref{eq14}) and (\ref{eq15})
    \begin{equation}
    \label{eq16}
    \begin{split}
    \mathop {\lim }\limits_{N\to \infty } P_{e\vert {\rm {\bf \uptheta }}} & = P\left[ {\frac{l(\uptheta)}{N} <\frac{E_s}{\beta }}\right]
    =P\left[ {l_n(\uptheta) <\frac{1}{\beta }}\right]\\
    & =
    \begin{cases}
      1 & \frac{1}{\beta }>l_n (\uptheta )\\
      0 & \text{else}
    \end{cases}
    \end{split}
    \end{equation}
    Substituting (\ref{eq16}) into (\ref{eq10}) will get (\ref{eq2261}).

\begin{figure}[htbp]
\centering
  \includegraphics[width=0.5\textwidth]{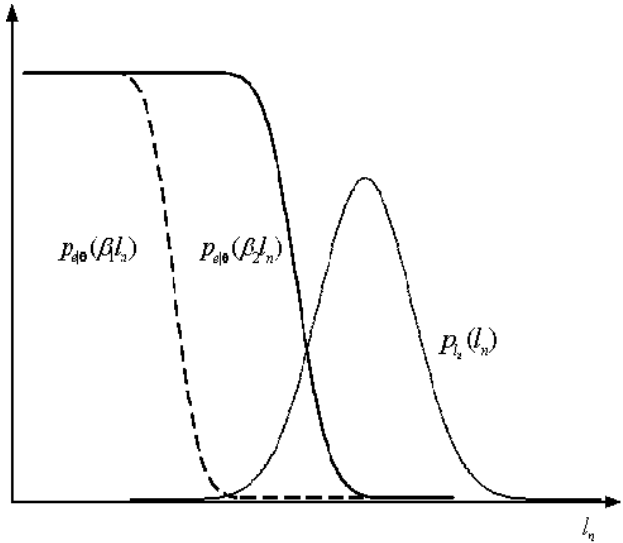}\\
  \caption{Illustration of $P_{e|\uptheta}(\beta l_n)$ and $p_{l_n}(l_n)$.($\beta_1>\beta_2$).}\label{f33}
\end{figure}

    The conditional WER (\ref{eq7}), i.e. the term in the square bracket of
    (\ref{eq10}) is a decreasing function around $l_n=1/\beta$ with respect to
    $l_n$. This is illustrated in Fig.\ref{f33} together with the SR-PDF $p_{l_n}(l_n)$. The WER of (\ref{eq10})
    is the integral of the product $P_{e|\theta}(\beta l_n)p_{l_n}(l_n)$. It is clear that when SNR is high,
    WER is dominated by the ¡°left tail¡± of the normalized SR-PDF.

    \begin{theorem}
      The \textit{inflection point} $\tau_0$ of $P_e^a(\tau)$, is the \textit{maximum point} of the normalized SR-PDF.
    \end{theorem}
    \begin{proof}
      Based on \emph{property 1}, taking the second derivative of (\ref{eq2261}) with respect to $\tau$
      \begin{equation}
      \label{eq22611}
      \frac{\text{d}^2P_e^a(\tau)}{\text{d}\tau^2}=\frac{\text{d} p_{l_n}(\tau)}{\text{d} \tau}=0
      \end{equation}
      The result $\tau_0$ that make (\ref{eq22611}) equal to zero is the inflection point of $P_e^a$ and
      must be the maximum point of $p_{l_n}(\tau)$.
    \end{proof}

\begin{figure}[htbp]
    \centering
    \includegraphics[width=0.5\textwidth]{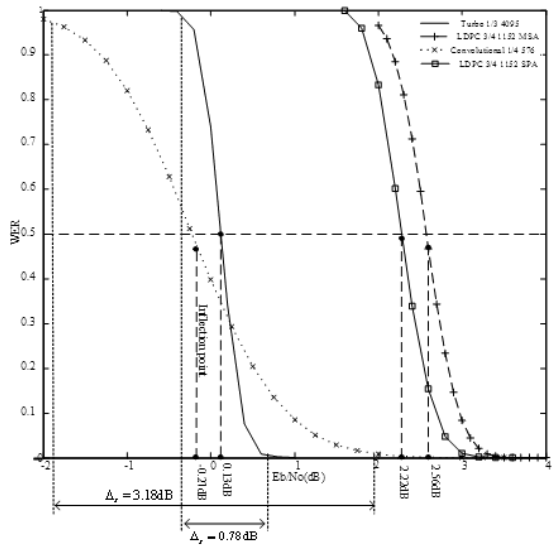}\\
    \caption{Critical SNR and $\Delta_{\epsilon}$. WER is the simulation result. Critical SNR is estimated by
    $1/\mu_{l_n}$. $\Delta_\epsilon$ is calculated from (\ref{Delta}).}\label{f32}
    \end{figure}

    The inflection point $\tau_0$ tells the position where the WER curve falls rapidly. Define \emph{critical SNR} as
    the inverse of the inflection point:
    \begin{equation}
      \label{eq2262}
      \beta_c=1/\tau_0
    \end{equation}
    then $\beta_c$ can be viewed as a single parameter which can be used to characterize the WER performance of long
    codes. This has been shown in Fig.~\ref{f32}, where we have drawn the simulated WER with liner coordinates.
    For the popular codes, the normalized SR-PDF tends to be symmetric about the maximum point and the
    normalized SR-PDF $p_{l_n}(x)$ arrives to its maximum roughly at $\mu_{l_n}=E[l_n]$.
    Thus, the critical SNR of a code can be approximated as $\beta_c\approx\frac{1}{\mu_{l_n}}$.
    The critical SNRs of the codes in Fig.~\ref{f3b} are listed in Table~\ref{tab3}.

    \begin{theorem}
    For the capacity achievable codes, decision region is a multi-dimensional sphere with constant radius.
    The inverse of the normalized square radius, i.e. the critical SNR $\beta_c$, is the Shannon limits.
    By "Shannnon limit", it means such a threshold SNR $\beta_s$ that for a given family of codes with fixed code rate
    and codeword length $N\rightarrow\infty$, if the channel SNR is greater than $\beta_s$, the code will be decoded
    successfully, otherwise, if SNR is less than $\beta_s$, the decoding process will fail.
    \end{theorem}
    \begin{proof}
    Based on \ref{eq10} and \ref{eq20}
        \begin{equation}
        \label{eq2263}
            P_e^a(\tau)=\int_0^{1/\beta}p_{l_n}(x)dx=
            \begin{cases}
              1\quad \beta\leq\beta_c\\
              0\quad \beta>\beta_c
            \end{cases}
        \end{equation}
    The $\beta_s$ that satisfy \ref{eq2263} is unique for the given family of codes, thus with the definition of $\beta_s$,
    it is clearly that $\beta_s=\beta_c$. \ref{eq2263} implies that $p_{l_n}(x)=\delta(x-\frac{1}{\beta_s})$, where $\delta(x)$ is the Dirac impulse
    function. A random variable with $\delta$ pdf is in fact a constant which is also the mean.
    Thus the decision region must be a sphere with constant radius.
    \end{proof}

\begin{figure}[htbp]
\begin{center}
\includegraphics[width=0.5\textwidth,keepaspectratio=true]{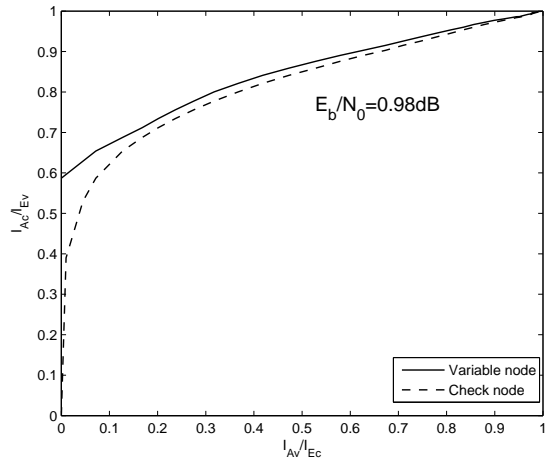}
\caption{EXIT chart for the LDPC code in Table~\ref{tab3}. The codeword length is 11520 and degree distribution
    is the same as the 1/2 rate LDPC code of Table~\ref{tab3}. $I_{A_v}$ and $I_{E_c}$ are the a priori and extrinsic mutual information
    between the transmitted signal and the soft information input to variable nodes and output from check nodes respectively.
    $I_{E_v}$ and $I_{A_c}$ are the extrinsic and a priori mutual information
    between the transmitted signal and the soft information output from variable nodes and input to check nodes respectively.}
\label{f07}
\end{center}
\end{figure}

    It is well known that threshold SNR of iterative soft-decision decoding can be obtained
    by means of Density Evolution \cite{16}\cite{17}
    or the equivalent (Extrinsic Information Transfer) EXIT chart \cite{18}\cite{19}\cite{20}.
    So property 3 implies that measuring the average square radius of the decision region is another way
    to determine threshold SNR. For the Turbo code in Table~\ref{tab3}, which is exactly the same as the one used in \cite{17},
    the threshold SNR obtained with Density Evolution can be found in \cite{17} as 0.70dB for 1/2 code rate and 0.02dB for 1/3 code rate,
    the difference with $\beta_c$ is within 0.07dB.
    For the 1/2 code rate LDPC code in Table~\ref{tab3}, its EXIT chart is shown in Fig.\ref{f07},
    where the two curves intersect until $I_{E_v}=1$ implying that the threshold SNR is about 0.98dB,
    while the critical SNR obtained from $\mu_{l_n}$ is 0.92dB, the difference is 0.06dB.

    Note that though the threshold SNR is generally viewed as a kind of \emph{analytical value},
    this value can only be obtained via numerical simulations with possible approximations (such as Gaussian approximation \cite{16}\cite{17}).
    Therefore, it is hard to say which one among $\beta_c$ and $\beta_s$ is more accurate.

    Property 3 has provided us a simple method to estimate the asymptotic performance (Shannon limit)
    for a given family of code. We only need to measure the mean of the square-radius with a code of adequate length
    because the mean is independent of codeword length if the code family is given.
    Note that measuring the mean is much simpler than measuring the pdf, with our experience, 1000 radius is enough
    to get a relatively accurate result.

    The difference between good codes (the codes that can achieve Shannon limits when $N\rightarrow\infty$)
    and practical codes is that, The WER of good codes falls steeply at the critical SNR while the WER of
    practical codes falls in a rolling off fashion around the critical SNR.
    Define $\Delta_{\epsilon}$ as the range corresponding to $P_e^a(\tau)$ falls from
    $1-\epsilon$ to $\epsilon$, where $0<\epsilon\ll1$, i.e.
    \begin{equation}\label{Delta}
    \Delta_\epsilon\triangleq\tau_{1-\epsilon}-\tau_{\epsilon}=F_{l_n}^{-1}(1-\epsilon)-F_{l_n}^{-1}(\epsilon)
    \end{equation}
    Then $\Delta_\epsilon$ can be viewed as a measure of perfectness
    of practical codes compared with the good codes.
    If the code is capacity achievable, the WER is a step function thus $\Delta_{\epsilon}=0$. Otherwise, $\Delta_{\epsilon}$
    will be a positive number. The $\Delta_\epsilon$ for codes in Fig.~\ref{f3}b are listed in table~\ref{tab3}.

    Assume that $p_{l_n}(\tau)$ is symmetrical about the mean $\mu_{l_n}$ when codeword length $N\rightarrow\infty$,
    $\Delta_\epsilon$ is bounded by
    \begin{equation}\label{Delta_bound}
    \Delta_\epsilon\leq\sqrt{2\sigma_{l_n}^2/\epsilon}
    \end{equation}
    where $\sigma_{l_n}^2$ is the variance of $l_n$. This is because that,
    with the \emph{Chebyshev Inequality}:
    \begin{equation}
      \label{eq32}
      2\epsilon=P\left\{|\tau-\tau_0|\geq\frac{\Delta_{\tau}}{2}\right\}\leq\frac{4\sigma_{l_n}^2}{\Delta_{\tau}^2}
    \end{equation}

    In practice, the WER falling range in decibel domain may be more interesting, i.e.
    \begin{equation}\label{Delta_db}
    \Delta_{\epsilon\text{(dB)}}\triangleq10\log_{10}\frac{\tau_2}{\tau_1}=
    10\log_{10}\frac{F_{l_n}^{-1}(1-\epsilon)}{F_{l_n}^{-1}(\epsilon)}
    \end{equation}
    The Chebyshev bound now changes to
    \begin{equation}\label{Delta_bound_db}
    \Delta_{\epsilon\text{(dB)}}\leq10\text{log}_{10}\left(\frac{\tau_0+\frac{\Delta_{\tau}}{2}}{\tau_0-\frac{\Delta_{\tau}}{2}}\right)
    =10\text{log}_{10}\left(\frac{\mu_{l_n}+\sigma_{l_n}\sqrt{1/2\epsilon}}{\mu_{l_n}-\sigma_{l_n}\sqrt{1/2\epsilon}}\right)\rm{dB}.
    \end{equation}
    The bounds in dB for codes in Fig.~\ref{f3}b are also listed in table~\ref{tab3}. It is obvious that the bound is quite
    loose, roughly $3\thicksim5$ times larger.

\section{Approximation of WER}
    Eq.(\ref{eq10}) involves an integration of the product of incomplete gamma function and SR-PDF. Moreover, accurate measurement of SR-PDF
    requires a large number of decoding tests. Thus, it is inconvenient for practical use, and approximation formulas will
    be welcome. The approximation of (\ref{eq10}) is to approximate the normalized SR-PDF $p_{l_n}(x)$. Although
    other approximations such as Gaussian are also possible, it is found that Gamma approximation is more accurate.
    The pdf of Gamma distribution is given by
    \begin{equation}
        \label{eq11}
        p(x)=\frac{1}{b^a\Gamma (a)}x^{a-1}e^{-\frac{x}{b}}
    \end{equation}
    where $a$ and $b$ are the parameters of Gamma distribution that have the following relationship with
    $\mu _{l_n} $ and $\sigma _{l_n}^2 $ \cite{15}:
    \begin{equation}
        \label{eq13}
        \begin{cases}
        a=\mu _{l_n}^2 /\sigma _{l_n}^2\\
        b=\sigma _{l_n}^2 /\mu _{l_n}
        \end{cases}
    \end{equation}
    Substitute (\ref{eq11}) into (\ref{eq10}),
    \begin{equation}
        \label{eq122}
        P_e \approx \int_{0}^\infty {\frac{1}{b^a\Gamma
            (a)}x^{a-1}e^{-\frac{x}{b}}\left[ {1-\frac{1}{\Gamma \left( {N
            \mathord{\left/ {\vphantom {N 2}} \right. \kern-\nulldelimiterspace} 2}
            \right)}\gamma \left( {\frac{N}{2},\frac{N\beta x}{2}} \right)} \right]dx}
    \end{equation}
    For long codes, increasing code length from $N$ to $N+1$ generally brings no notable performance
    difference. So assuming that $N$ is even and recalling (\ref{eq9}), the WER can be simplified as
\begin{equation}
  \label{eq12}
  \begin{split}
    P_e &\approx \sum\limits_{m=0}^{\frac{N}{2}-1} {\frac{\left(\frac{N\beta}{2}\right)^m}{b^a\Gamma (a)m!}\int_0^\infty
        {x^{m+a-1}e^{-x(\frac{1}{b}+\frac{N\beta x}{2})}} } dx \\
        &=\sum\limits_{m=0}^{\frac{N}{2}-1} {\frac{\left(\frac{N\beta}{2}\right)^m\left( {\frac{2b}{2+N\beta b}}
        \right)^{m+a}\Gamma (m+a)}{b^a\Gamma(a)\Gamma(m+1)}\int_0^\infty \frac{1}{\left( {\frac{2b}{2+N\beta b}} \right)^{m+a}\Gamma
        (m+a)} {x^{m+a-1}e^{-\frac{x(2+N\beta b)}{2b}}} } dx \\
        &=\sum\limits_{m=0}^{\frac{N}{2}-1}\frac{\left( 1-u \right)^mu^a}{mB(a,m)}  \\
  \end{split}
\end{equation}
    where $B(a,m)=\frac{\Gamma(a)\Gamma(m)}{\Gamma(a+m)}$ is the \textit{Beta Function} \cite{12} and $u={\frac{2}{N\beta b+2}}<1$.
    The advantage of (\ref{eq12}) over (\ref{eq10}) is that only $\mu _{l_n}$ and $\sigma _{l_n}^2 $ need to be measured.
    In the viewpoint of statistics,
    the number of radiuses required to estimate $\mu _{l_n} \text{ and } \sigma _{l_n}^2 $ is much smaller than to estimate the SR-PDF.
    Table~\ref{tab1} lists the mean and variance of $l_n$ and $a,b$ for some codes in AWGN channel.

\begin{figure}[htbp]
    \begin{center}
    \includegraphics[width=0.5\textwidth,keepaspectratio=true]{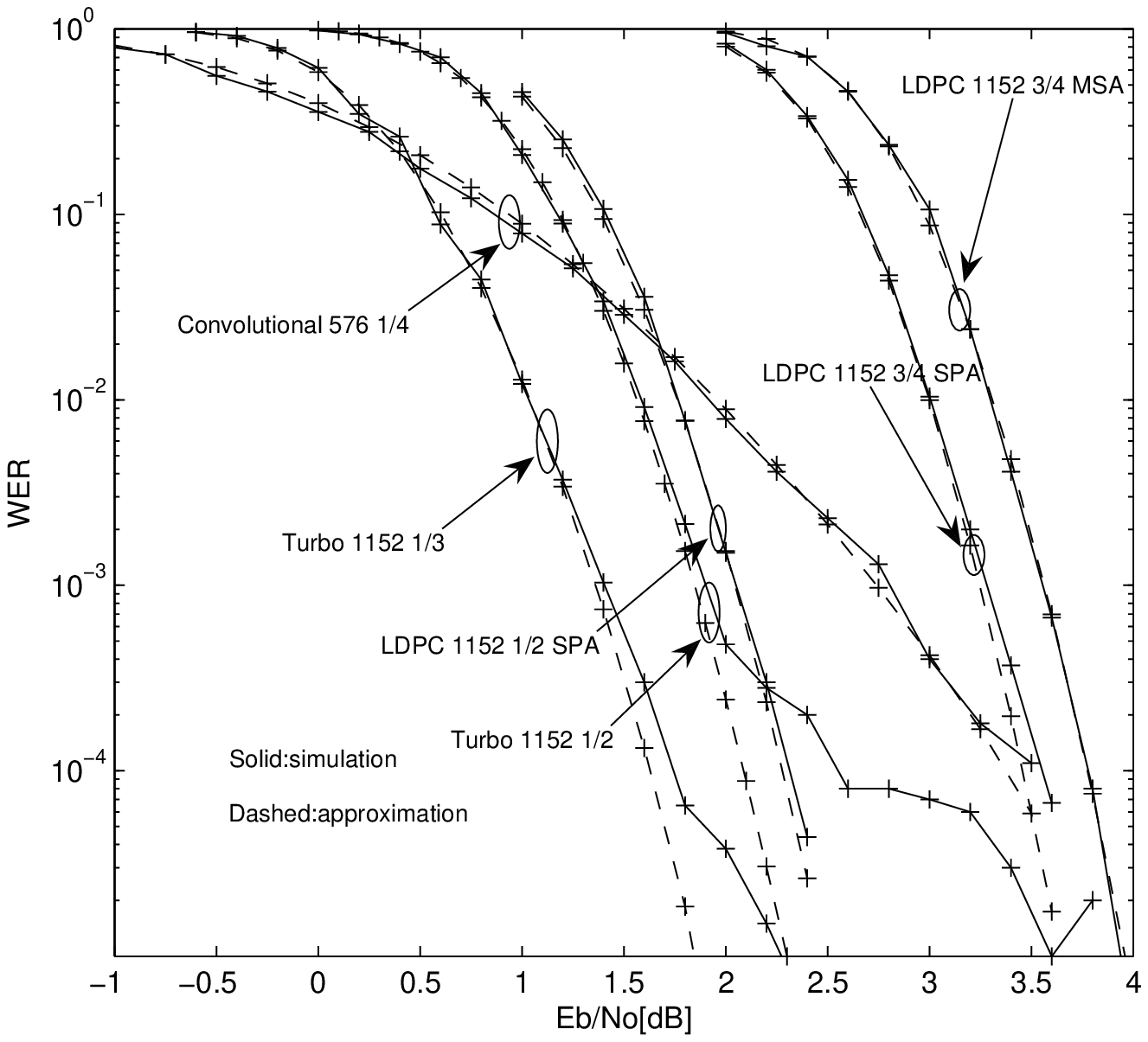}
    \caption{Comparison of WER between simulation and approximation using (\ref{eq12}) for
    Convolutional codes, Turbo codes and LDPC codes in AWGN channel.
    The decoding algorithms are soft decision Viterbi decoding for Convolutional codes,
    Sum-Product algorithm (SPA) and Min-Sum algorithm (MSA) respectively with 25 maximum iterations for LDPC codes
    and Log-MAP with 8 maximum iterations for Turbo codes.}
    \label{f4}
    \end{center}
\end{figure}

    Fig.~\ref{f4} is the comparison of WER between simulation and
    approximation using (\ref{eq12}). It can be seen that the approximation
    for all the codes only deviates from the simulation within about 0.1dB
    for WER above $10^{-3}$. If the error floor of approximated Turbo code WER
    does not occur at high $E_b/N_0$, this is a nice approximation for WER of interest with
    a significantly lower computational complexity.

    For a large codeword length, the WER expression can be further simplified using \emph{property 1} in section III.
    Substitute (\ref{eq11}) into (\ref{eq2261}), the WER can be approximated as
    \begin{equation}
    \label{eq18}
    \setlength\jot{10pt}
    \begin{split}
      P_e & \approx \int_{0 }^{1/\beta } {\frac{1}{b^a\Gamma(a)}x^{a-1}e^{-\frac{x}{b}}} dx \\
        & =\frac{1}{\Gamma (a)} \gamma \left(a,\frac{1}{\beta b}\right) \\
    \end{split}
    \end{equation}

\begin{figure}[htbp]
    \begin{center}
    \includegraphics[width=0.5\textwidth,keepaspectratio=true]{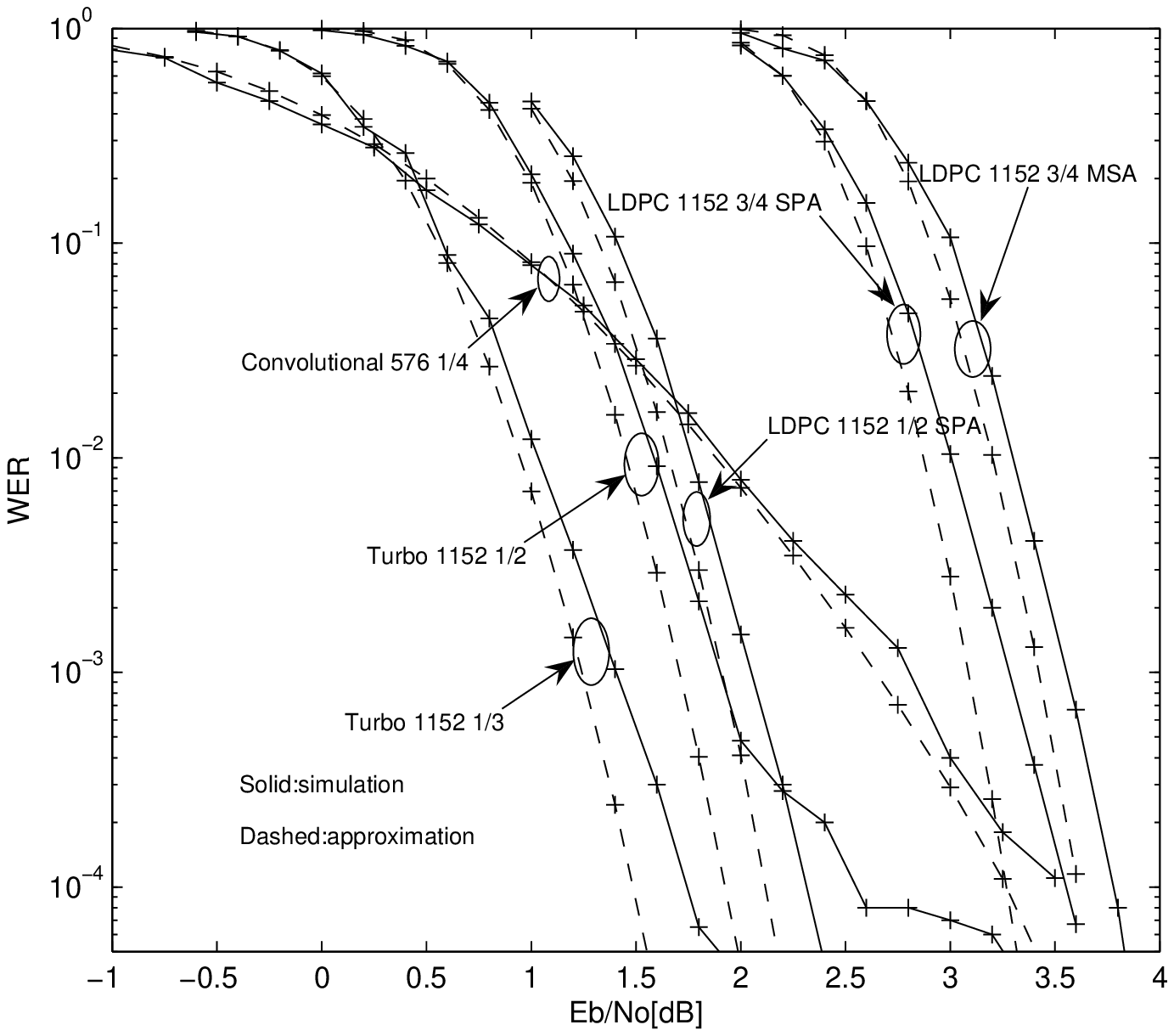}
    \caption{Comparison of WER between simulation and approximation using (\ref{eq19}) for
    Convolutional codes, Turbo codes and LDPC codes in AWGN channel.
    The decoding algorithms are soft decision Viterbi decoding for Convolutional codes,
    Sum-Product algorithm (SPA)
    and Min-Sum algorithm (MSA) respectively with 25 maximum iterations for LDPC codes and
    Log-MAP with 8 maximum iterations for Turbo code.}
    \label{f5}
    \end{center}
    \end{figure}

    In (\ref{eq18}), rounding $a$ to its lower integer and recalling (\ref{eq9}), the WER can be approximated as
    \begin{equation}
    \label{eq19}
      P_e \approx 1- e^{-\frac{1}{\beta b}} \sum_{m=0}^{\left\lfloor a \right\rfloor -1} \frac{\left(\frac{1}{\beta b}\right)^m}{m!} \\
    \end{equation}
    (\ref{eq19}) and (\ref{eq12}) use the same parameters to approximate WER.
    (\ref{eq19}) is simpler to evaluate but less accurate than (\ref{eq12}).
    Fig.~\ref{f5} is the comparison between simulation WER and approximation of (\ref{eq19}).
    Similar to the result of Fig.~\ref{f4}, If the error floor of Turbo code does not occur
    within the range that WER is above $10^{-3}$, the approximation result deviate from the simulation within about 0.3dB.

    When (\ref{eq12}) and (\ref{eq19}) are applied to evaluate the WER, an important problem is how accurate should the parameters $a$ and $b$
    be measured.
    Generally, for a long code $a$ possesses a large value. It can be verified that the term $\frac{u^a}{B(a,m)}$ in (\ref{eq12})
    is insensitive to the error of $a$, and the last few terms in the summation of (\ref{eq19}) are very small
    (on the order of two magnitudes lower than the sum).
    Thus an error of $a$ within $\pm1$ is safely acceptable for the WER evaluation.
    The parameter $b$ only occurs in (\ref{eq19}) in the form of product $\beta b$. Thus, an error of $b$ by $\Delta$dB is equivalent to
    that $b$ is exact but $\beta$ (SNR) is biased by $-\Delta$dB. From Fig.~\ref{f4} and Fig.~\ref{f5}, it can be observed that
    the accuracies of (\ref{eq12}) and (\ref{eq19}) are on the order of 0.1dB and 0.3dB respectively,
    therefore, the measurement error of $b$ should be less than 0.1dB. Several hundreds of radiuses are generally enough for this requirement.

\section{Word Error Rate In Flat Fading Channel}
    In flat fading channel, (\ref{eq1}) changes to
    \begin{equation}
    \label{eq20}
    {\rm {\bf y}}=diag(\bf h)\cdot {\rm {\bf s}}+{\rm {\bf n}}
    \end{equation}
    where ${\rm {\bf h}}=\left[ {h_1 ,h_2 ,...,h_N } \right]^T$ is the channel gain, which scales every element of the transmitted vector.
    Thus, the decision region is dependent on a specific channel vector. Given the channel vector,
    the conditional WER can be calculated by (\ref{eq10}) with $p_{l_n}(l_n)$ be replaced by $p_{l_n|\bf h}(l_n)$
    which is the normalized SR-PDF conditioned on a channel vector. The average WER for flat fading channel is then obtained
    by taking expectation over all possible $\bf h$:
    \begin{equation}
    \label{eq21}
    \setlength\jot{8pt}
        \begin{split}
        \overline{P_e} & =E_{\rm
            {\bf h}} \left\{ {\int_0^\infty {\left[ {1-\frac{1}{\Gamma \left( {N
            \mathord{\left/ {\vphantom {N 2}} \right. \kern-\nulldelimiterspace} 2}
            \right)}\gamma \left( {\frac{N}{2},\frac{N\beta x}{2}} \right)}
            \right]p_{l_n\vert {\rm {\bf h}}} (x)} dx} \right\} \\
            &=\int_0^\infty {\left[ {1-\frac{1}{\Gamma \left( {N \mathord{\left/
            {\vphantom {N 2}} \right. \kern-\nulldelimiterspace} 2} \right)}\gamma
            \left( {\frac{N}{2},\frac{N\beta x}{2}} \right)} \right]E_{\rm {\bf h}}
            \left[ {p_{l_n\vert {\rm {\bf h}}} (x)} \right]} dx \\
            &=\int_0^\infty {\left[ {1-\frac{1}{\Gamma \left( {N \mathord{\left/
            {\vphantom {N 2}} \right. \kern-\nulldelimiterspace} 2} \right)}\gamma
            \left( {\frac{N}{2},\frac{N\beta x}{2}} \right)} \right]\bar {p}_{l_n} \left( x
            \right)} dx \\
        \end{split}
    \end{equation}
    where $\bar {p}_{l_n}(x)\triangleq E_{\bf h}[p_{l_n|\bf{h}}(x)]$ is the normalized SR-PDF averaged over the ensemble
    of ${\rm {\bf h}}$.

\begin{figure}[htbp]
\begin{center}
\includegraphics[width=0.5\textwidth,keepaspectratio=true]{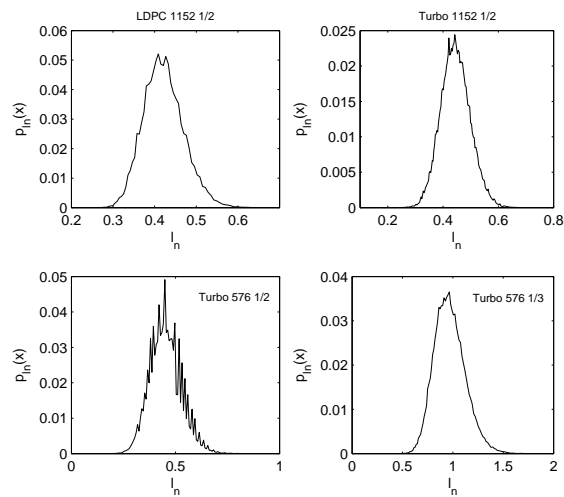}
\caption{Normalized average SR-PDF $\bar {p}_{l_n}(x)$ for several
representative error control codes in Rayleigh flat fading channel.}
\label{f8}
\end{center}
\end{figure}

    Fig.~\ref{f8} is the normalized average SR-PDF for several codes investigated in this
    paper. In these examples, fully interleaved Rayleigh flat fading channel is
    considered. The elements of ${\rm {\bf h}}$ are i.i.d. Rayleigh random variables with pdf
    \begin{equation}
    \label{eq22}
    p_h (h_i )=2h_i e^{-h_i^2 },~h_i >0, ~i=1,2,\cdots,N
    \end{equation}
    Fig.~\ref{f8} indicates that the average SR-PDF in fully interleaved Rayleigh flat fading channel still keeps the same shape as in AWGN channel. Therefore,
    the approximations presented in Section IV, i.e. (\ref{eq12}) and (\ref{eq19}) can also be used to evaluate the average WER
    in flat fading channels,
    only with $a$ and $b$ replaced by $\overline{a}$ and $\overline{b}$ respectively, which are averaged over all channel gain realizations.
    Table II lists the mean and variance of the average $l_n$ and $\overline{a},\overline{b}$ used in (\ref{eq12}) and (\ref{eq19}) in
    Rayleigh flat fading channel.

\begin{figure}[htbp]
\begin{center}
\includegraphics[width=0.5\textwidth,keepaspectratio=true]{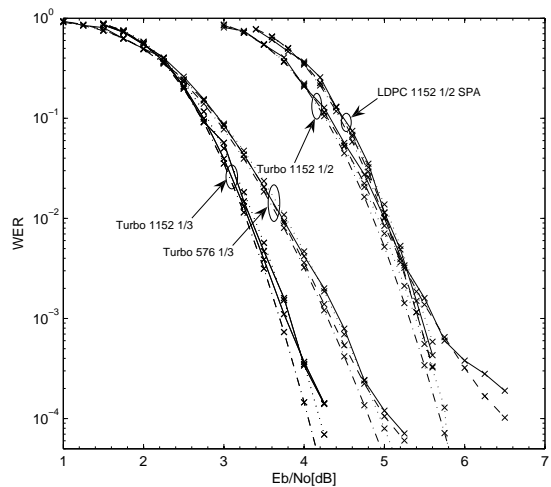}
\caption{Comparison of WER between simulation and approximations of
Turbo codes, LDPC codes in Rayleigh flat fading channel. The
decoding algorithms are Log-MAP with 8 maximum
iterations for Turbo codes and SPA with 25 maximum iterations for LDPC codes.
Solid: simulation; Dashed: approximation using
(\ref{eq21}); Dash dot: approximation using (\ref{eq19}); Dotted:
approximation using (\ref{eq12})} \label{f9}
\end{center}
\end{figure}

    Fig.~\ref{f9} is the comparison between simulation WER and (\ref{eq21}), (\ref{eq12}), (\ref{eq19}).
    It is obvious that the SR-PDF method and its approximations are still applicable in flat fading channel.

\section{Conclusion}
    SR-PDF of decision region introduces a new method to evaluate the
    performance of binary block codes. The WER can be calculated using this pdf
    precisely, and even the closed-form approximations are more precise than
    existing tightest bounds for practically used long block codes at SNRs of interesting.
    Despite that the SR-PDF method is demonstrated with binary codes in AWGN and flat fading channel in this paper,
    it is straightforward to generalize this method to any situations where the error rate is characterized by the decision region,
    such as memoryless modulation, MIMO detection, coded-modulation, equalization, etc.
    In these situations, the decision region may not have the same shape for different transmitted signals.
    Nevertheless, the average error rate can still be evaluated by the average SR-PDF in a similar way as in fading channel.

\newpage
    \begin{table}[htbp]
    \centering
    \caption{\label{tab3} The Critical SNR and $\Delta_{\epsilon}$ ($\epsilon$=0.01) for some codes.}
    \begin{tabular}{|l|l|l|l|}
      \hline\hline
      Codes & Critical SNR ($E_b/N_{0_{dB}}$) & $\Delta_{\epsilon\text{(dB)}}$ & upper bound of $\Delta_{\epsilon\text{(dB)}}$\\
      \hline
      Turbo 1152 1/3 (Log-MAP) & 0.09 & 1.65 & 5.80\\
      \hline
      Turbo 576 1/3 (Log-MAP) & 0.03 & 2.12& 9.88\\
      \hline
      Turbo 1152 1/2 (Log-MAP) & 0.76 & 1.64 & 4.54 \\
      \hline
      LDPC 1152 1/2 (SPA) & 0.92 & 1.15 & 4.46\\
      \hline
      LDPC 1152 3/4 (SPA) & 2.22 &1.21 & 3.73 \\
      \hline\hline
    \end{tabular}
    \begin{tablenotes}
    \centering
      \item [1] * The parameters $\mu_{l_n}$ and $\sigma_{l_n}^2$ used to calculate the critical SNR and the upper bound
      are listed in table~\ref{tab1}.
    \end{tablenotes}
    \end{table}
    \begin{table*}[htbp]
    \renewcommand{\arraystretch}{0.8}
    \centering
    \caption{\label{tab1}Parameters of normalized SR-PDF for representative error control codes in AWGN channel}
        \begin{tabular}{|l|l|l|l|l|l|}
            \hline \hline
            codes&
            decoding
            algorithm&
            mean&
            variance&
            a&
            b \\
            \hline
            Turbo 1152 1/3&
            Log-MAP&
            1.47&
            1.47e-2&
            147.45&
            1.00e-3 \\
            \hline
            Turbo 1152 1/2&
            Log-MAP&
            0.84&
            3.25e-3&
            219.55&
            3.85e-3 \\
            \hline
            Turbo 1152 1/2&
            Max-Log-MAP&
            0.79&
            3.06e-3&
            202.18&
            3.89e-3 \\
            \hline
            Turbo 576 1/3&
            Log-MAP&
            1.49&
            2.94e-2&
            75.39&
            1.98e-2 \\
            \hline
            Turbo 576 1/3&
            Max-Log-MAP&
            1.39&
            2.81e-2&
            68.21&
            2.03e-2 \\
            \hline
            Turbo 576 1/2&
            Log-MAP&
            0.85&
            6.30e-3&
            115.23&
            7.40e-3 \\
            \hline
            Turbo 1152 2/3&
            Log-MAP&
            0.51&
            1.13e-3&
            233.37&
            2.20e-3 \\
            \hline
            Turbo 1152 3/4&
            Log-MAP&
            0.40&
            7.84e-4&
            204.04&
            1.96e-3 \\
            \hline
            LDPC 1152 1/2&
            Sum-Product&
            0.81&
            2.93e-3&
            221.19&
            3.64e-3 \\
            \hline
            LDPC 1152 1/2&
            Min-Sum&
            0.72&
            2.13e-3&
            245.43&
            2.95e-3 \\
            \hline
            LDPC 1152 3/4&
            Sum-Product&
            0.40&
            5.25e-4&
            298.66&
            1.33e-3 \\
            \hline
            LDPC 1152 3/4&
            Mean-Sum&
            0.37&
            4.76e-4&
            286.14&
            1.29e-3 \\
            \hline
            LDPC 1152 2/3&
            Sum-Product&
            0.50&
            8.89e-3&
            280.85&
            1.78e-3 \\
            \hline
            Convolution 576 1/4&
            Viterbi-Soft&
            2.13&
            1.68e-1&
            27.08&
            7.88e-2 \\
            \hline
            Convolution 576 1/3&
            Viterbi-Soft&
            1.45&
            6.01e-2&
            34.85&
            4.15e-2 \\
            \hline
            Convolution 576 1/2&
            Viterbi-Soft&
            0.82&
            1.45e-2&
            46.68&
            1.76e-2 \\
            \hline \hline
        \end{tabular}
    \end{table*}
\begin{table}[htbp]
\begin{center}
\caption{\label{tab2}Parameters of average normalized SR-PDF for
representative error control codes in Rayleigh~flat~fading~channel}
    \begin{tabular}{|l|l|l|l|l|l|}
        \hline \hline
        codes&
        decoding
        algorithm&
        mean&
        variance&
        a&
        b \\
        \hline
        Turbo 1152 1/3&
        Log-MAP&
        0.935&
        1.12e-2&
        77.70&
        1.20e-2\\
        \hline
        Turbo 1152 1/2&
        Log-MAP&
        0.442&
        2.89e-3&
        67.59&
        6.54e-3 \\
        \hline
        Turbo 576 1/3&
        Log-MAP&
        0.958&
        2.30e-2&
        39.70&
        2.41e-2 \\
        \hline
        LDPC 1152 1/2&
        Sum-Product&
        0.419&
        2.27e-3&
        77.18&
        5.43e-3 \\
        \hline \hline
    \end{tabular}
\end{center}
\end{table}

\end{document}